# Using Electrical Impedance Spectroscopy to Separately Quantify the Effect of Strain on Nanosheet and Junction Resistance in Printed Nanosheet Networks

Eoin Caffrey[1], Tian Carey[1], Luke Doolan[1], Anthony Dawson[1], Emmet Coleman[1], Zdeněk Sofer[2], Oran Cassidy[1], Cian Gabbett[1], Jonathan N. Coleman[1]*

[1]*School of Physics, CRANN & AMBER Research Centres, Trinity College Dublin, Dublin 2, Ireland*

[2]*Department of Inorganic Chemistry, University of Chemistry and Technology Prague, Technická 5, 166 28 Prague 6, Czech Republic*

*colemaj@tcd.ie (Jonathan N. Coleman); Tel: +353 (0) 1 8963859.

ABSTRACT: Many printed electronic applications require strain-independent electrical properties to ensure deformation-independent performance. Thus, developing printed, flexible devices using 2D and other nanomaterials will require an understanding of the effect of strain on the electrical properties of nano-networks. Here we introduce novel AC electrical techniques to fully characterise the effect of strain on the resistance of high mobility printed networks, fabricated from of electrochemically exfoliated $MoS_2$ nanosheets. These devices were initially characterised using DC piezoresistance measurements and showed good cyclability and a linear strain response, consistent with a low gauge factor of G~3. However, AC impedance spectroscopy measurements, performed as a function of strain, allowed the measurement of the effects of strain on both the nanosheets and the inter-nanosheet junctions separately. The junction resistance was found to increase linearly with strain, while the nanosheet resistance remained constant. This response is consistent with strain-induced sliding of the highly-aligned nanosheets past one another, without any strain being transferred to the sheets themselves. Our approach allows us to individually estimate the contributions of dimensional factors (G~1.4) and intrinsic factors (G~1.9) to the total gauge factor. This novel technique may provide insight into other piezoresistive systems.

INTRODUCTION

Piezoresistance is a physical phenomenon which has been observed in a range of materials, from metal foils[1] to bulk semiconductors[2] and more recently networks and composites of nanomaterials.[3] It manifests itself as a change in the electrical resistance with applied strain. The parameter which quantifies the magnitude of this effect is the gauge factor, G, which is defined as the fractional resistance change ($\Delta R/R_0$) per unit strain (ε) in the linear, low-strain regime, $\Delta R/R_0 = G\varepsilon$.[1, 4] Here, $R_0$ denotes the resistance in the unstrained state. Usually, the gauge factor is measured at low strain and in that limit it can be shown that for an isotropic material[5]

$$G = (1+2\nu) + \frac{1}{\rho_0}\frac{d\rho}{d\varepsilon} \qquad (1)$$

where $\nu$ and $\rho$ are the Poisson ratio and resistivity of the piezoresistive material and $\rho_0$ is its zero-strain resistivity. The first term describes the effect of strain on sample dimensions (and so resistance) while the second term describes the effect of strain on the intrinsic material properties. Much work has been done to understand piezoresistance in a broad range of materials.[3, 4, 6-12] From an applications standpoint, a major focus has been to develop piezoresistive materials with high gauge factors which can be used to fabricate extremely sensitive strain and pressure sensors.[3, 13]

However, understanding the origin of piezoresistive effects in low gauge factor systems is also extremely important. Printed networks of nanoparticles such as 2D nanosheets or 1D nanowires are important for a range of applications including transparent conductors,[14, 15] transistors,[16-20] solar cells[21] and photodiodes.[22, 23] One significant advantage of such systems is that they have great potential to be used in flexible and stretchable devices.[24-26] One key aspect for such devices is that their electrical properties should be stable under mechanical deformation, i.e. low gauge factors are required. For example, for a device's resistance to change by <5% under an applied strain of 1%, it must have G<5. Printed networks of 2D nanosheets are important for a range of flexible electronic applications. It has been shown that printed transistors fabricated from networks of $MoS_2$ nanosheets, yield high performance transistors[18] which retain good performance under mechanical deformation.[16] In addition, self-healing graphene oxide hydrogel structures have been produced with gauge factors as low as 0.01 for use in implantable bioelectronics.[27] It will be important to understand the piezoresistance of such systems, not to maximise gauge factor but to minimise it.

Traditionally, the interpretation of piezoresistive data from nanocomposite and nanoparticle networks has focused on the effects of strain on the resistance associated with charge transport across inter-particle junctions, i.e. the junction resistance.[6, 28-30] It is widely believed that increasing the applied strain results in an increase in the mean junction resistance between conductive particles,[8, 31, 32] although other factors such as the intrinsic piezoresistive response of the nanoparticles[33, 34] and the effect of deformation on the overall network structure are also present.[35-37] Indeed, although strain-induced changes in junction resistance certainly play an important role in network piezoresistance, we have found no examples of direct measurements of the junction resistance while the material is strained. Such a direct measurement would be extremely interesting as it would significantly enhance our understanding of piezoresistance mechanism. This is largely because such an in-situ measurement of mean junction resistance in the presence of an applied strain is extremely challenging. However, very recently a technique based on electrical impedance spectroscopy[38] has been developed wherein AC electrical measurements on a printed nanosheet network allow both the mean junction resistance and the mean resistance of individual nanosheets to be measured.

In this paper, we show that this impedance technique can be used to study the piezoresistance of a model nanostructured system consisting of a printed network of electrochemically exfoliated $MoS_2$ nanosheets. Such networks are technologically important for use as high-mobility printed transistors.[16-18, 39] In such applications, it is important that their gauge factors are low to minimize variations in electrical performance under deformation.[40] This makes this important to understand their piezoresistive properties. The novel impedance technique utilized here allows us to separately measure the effects of strain on junction resistance and nanosheet resistance. In this way we show that, for this system at least, applying strain results in an increase in junction resistance but no measurable increase in the resistance of the nanosheets making up the network.

RESULTS AND DISCUSSION

*Material Characterisation*

Here we will use networks of electrochemically exfoliated $MoS_2$ nanosheets as a model system for the characterisation of network piezoresistance using impedance spectroscopy. The first step in this process is to produce and characterise the nanosheets which will later be used to fabricate the networks. The nanosheets used in this work were prepared by electrochemical exfoliation from bulk layered crystals of $MoS_2$ as described in previous publications.[16, 38] This procedure results in inks consisting of $MoS_2$ nanosheets dispersed in IPA. The nanosheet length and thickness for an identical ink to the one used here was measured using AFM, as reported by Gabbett *et al.*[38] to be $l_{NS}$ =1 µm and $t_{NS}$ =3.3 nm (see Table S1). Figure 1A shows a photograph of such an ink which shows the characteristic green tint associated with $MoS_2$ dispersions. The absorbance spectrum of the ink is shown in Figure 1B and confirms the presence of $MoS_2$, via the characteristic A and B exciton peaks[41] at 670 and 610 nm respectively. Raman spectra of the nanosheets (Figure 1C), measured on a thin film drop cast from the ink onto $Si/SiO_2$, show the characteristic $A_{1g}$ and $E^1_{2g}$ peaks associated with $MoS_2$ nanosheets.[42, 43]

We used these nanosheets to fabricate thin networks using a modified Langmuir-Schaeffer deposition method that has previously be used to prepare high-mobility transistors.[16, 38] This method involves injecting the nanosheet ink at the interface between two immiscible liquids, water and hexane.[44] The surface energy difference then drives the formation of a thin, highly aligned network of nanosheets at the interface. Lifting a substrate through the interface then results in the deposition of the network onto the substrate. This process can be repeated multiple times in order to build up thicker films. More information is given in Methods.

The morphology of the nanosheet networks can most easily be analysed via scanning electron microscopy (SEM) imaging of the top surface. Figure 1D shows a top-down view of the network of nanosheets fabricated on Kapton substrates using the Langmuir-Schaefer method. This image show highly-aligned nanosheets which overlap their neighbours via large area conformal junctions. This is typical for networks of high aspect ratio nanosheets such as those produced by electrochemical exfoliation.[16, 38] Such nanosheets combine considerable flexibility (because they are thin) with the ability to van der Waals bond strongly to their neighbours (because they have large area) enabling them to form large area, conformal junctions.[45] Such junctions display low junction resistance[38] leading to networks with high carrier mobility.[18, 46] For example, Carey *et al.* reported transistors based on electrochemically exfoliated, transition metal dichalcogenide nanosheets deposited using the Langmuir-Schaefer method with mobilities >10 $cm^2$/Vs.[16]

We can quantify the area of these conformal junctions from a series of SEM images. By estimating the overlap area of over 180 conformal inter-nanosheet junctions, we could generate a histogram of junction areas as shown in Figure 1E. This is consistent with a roughly log-normal distribution, with a mean junction overlap area of $A_{J,0}=0.33 \pm 0.03$ µm$^2$. This is broadly in line with the nanosheet overlap area measured by Gabbett *et al.* for similar MoS$_2$ films.[38] This result will allow us to analyse more deeply the properties of inter-nanosheet junctions within the networks.

*Basic Electromechanical properties (Linear Strain and Cycling)*

In order to perform electrical and piezoresistive measurements, we used the Langmuir-Schaefer method to deposit networks of electrochemically exfoliated MoS$_2$ nanosheets onto pre-deposited gold electrodes on Kapton substrates. The electrodes had a separation (channel length) of $L_{Ch}$=100 µm and a channel width of $W_{Ch}$=19.55 mm, while the network had a mean thickness of $t_{Net}$=(19.6 ± 1.2) nm. We found the zero-strain resistance of such a network to be R$_0$= 1.38 kΩ, consistent with a network conductivity of σ$_{net}$=190 S/m. This is somewhat larger than the value of 41 S/m reported by Gabbett *et al.*[38] probably due to differences in film thickness[35] and /or the effects of the Kapton substrate.

When studying the piezoresistance in any system, the first task is to investigate the material's response to quasi-static strain and to identify the regime in which the device's change in resistance is linear with strain.[1, 7] In our case, this was performed by mounting a nanosheet network deposited on a Kapton substrate in a modified tensile tester and straining at a rate of 0.05 %/s between 0% and 1.6% strain, while simultaneously measuring the resistance using a source meter. The resultant fractional change in resistance, $\Delta R / R_0$, is plotted vs the applied strain, $\varepsilon$, in Figure 2A. The gauge factor is defined as the slope of this plot in the low strain, linear regime. For this MoS$_2$ network, the gauge factor is measured to be G$_{Net}$=2.25, with a linear range of up to 1.2% strain.

As a comparison to similar large aspect ratio nanosheets, networks of electrochemically exfoliated graphene networks have reported gauge factors of 6, with a linear range of 1.5%.[47] These gauge factors are lower than for liquid phase exfoliated (LPE) nanosheets which show gauge factors of up to 350.[35] Although they did not report a gauge factor, our linear range is consistent with the observations of Chu *et al.* who studied MoS$_2$ nanosheets coated on carbon fibres and had a maximum linear range of 0.8%.[48] Our gauge factor is relatively low which means that the resistance of our networks is reasonably strain-invariant.

For any electrical material, an important consideration is the repeatability and cyclic behaviour of the piezoresistive response. This can be investigated through repeated cyclic straining using a triangular sawtooth pattern, while simultaneously measuring the resistance of the network. To keep the cyclic testing within the linear piezoresistive range, the strain was cycled between 0 and 0.5 %, with a strain rate of 0.2 %/s in the linear region, as shown in figure 2B. This data shows a well-defined resistance response which is sawtooth in nature and in phase with the applied strain.

From Figure 2B, the minimum ($R_{n,0}$) and maximum ($R_{n,\varepsilon_{Max}}$) resistance values for each cycle number, $n$, corresponding to the unstrained and maximum strain (0.5%) situations can be measured. From such data, the gauge factor for each cycle can be determined using the following equation.

$$G_n \approx \left( \frac{R_{n,\varepsilon_{Max}} - R_{n,0}}{R_{n,0}\varepsilon_{max}} \right) \qquad (2)$$

We have calculated this gauge factor for >50 cycles and plotted the results as a histogram which is displayed in figure 2C. We find the gauge factor to be quite stable, displaying values in the range 2-3 (mean 2.4 ± 0.2), consistent with the gauge factor extracted from the plot of linear strain.

This data clearly shows that the gauge factor of our highly aligned $MoS_2$ network is relatively low. To properly analyse the data, it will be necessary to understand the factors controlling conduction in networks of electrochemically exfoliated nanosheets such as those under study here. Recently Gabbett et al proposed a model which well-describes conduction in nanosheet networks.[38] This paper gives an equation for network resistivity which, in the case of very low-porosity networks of doped[38] nanosheets as we have here, can be written as (see SI of ref[38]):

$$\rho_{Net} \approx 2t_{NS}\left[R_{NS} + R_J\right] \qquad (3a)$$

where $t_{NS}$ is the mean nanosheet thickness and $R_{NS}$ and $R_J$ are the mean resistances of individual nanosheets and junctions within the network. Combining equation 1 and equation 3a allows us to generate an equation for the gauge factor of our networks. In addition to calculating the second term in equation 1, we modify the first term to take into account the fact that nanosheet networks are anisotropic and are expected to have different Poisson ratios in the in-plane (y) and out-of-plane (z) direction (see SI)

$$G_{Net} = (1 + \nu_y + \nu_z) + \frac{dR_{NS}/d\varepsilon + dR_J/d\varepsilon}{R_{NS,0} + R_{J,0}} \qquad (3b)$$

Here the subscript denotes the value at ε=0. The first term describes the contribution to the piezoresistance from dimensional changes under deformation ($G_{Dimensional}$) while the second term describes the intrinsic piezoresistance of the material ($G_{Intrinsic}$).

Equation 3b is useful as it shows that, even in the absence of intrinsic piezoresistance, networks will display a minimum gauge factor of $G_{Net} = (1 + \nu_y + \nu_z)$, simply due to dimensional changes upon deformation. What is unclear is whether such purely geometric effects can explain the observed gauge factor or whether there is a small intrinsic piezoresistance in these systems. More detailed measurements and analysis are required to obtain values of $R_{NS}$ and $R_J$ and their strain derivatives in order to understand the importance of any intrinsic piezoresistance.

*Impedance Spectroscopy of a network at different strain magnitudes.*

Impedance spectroscopy is a technique where the electrical response of a material is measured as a function of the frequency of an applied AC voltage. This technique has the advantage that it allows one to extract a broader range of data compared to DC measurements. Impedance spectra are often analyzed via comparisons to equivalent circuits. As discussed in detail in the SI of ref[38], for a nanosheet network, the relevant equivalent circuit is that shown in figure 3A (top). Here $R_S$ and $R_P$ represent the contribution to the network resistance of the resistances of all the nanosheets and all the inter-nanosheet junctions within the sample, respectively. Similarly, $C_P$ represents the sample capacitance which is governed by the capacitances of all the inter-nanosheet junctions within the sample. The microscopic origin of this equivalent circuit is the fact that the basic unit of conduction in a nanosheet network is the nanosheet-junction pair[38] which has a similar equivalent circuit (figure 3A (bottom). Here $R_{NS}$ and $R_J$ represent the mean resistance of individual nanosheets and junctions within the network while $C_J$ is the mean junction capacitance. Importantly, Gabbett et al recently demonstrated how to calculate $R_{NS}$, $R_J$ and $C_J$ from $R_S$, $R_P$ and $C_P$.[38]

The real (ReZ$_{Net}$) and imaginary (-ImZ$_{Net}$) impedance spectra, as measured for our MoS$_2$ networks at zero strain, are plotted in figure 3B and C respectively as a function of the angular frequency of the alternating voltage. Although the measurement was performed between 125 and $2\times10^8$ rad/s, only the upper end of this range is shown (the region between 125 and $10^6$ rad/s was featureless). The ReZ$_{Net}$ spectrum shows a plateau at low frequency followed by a rapid fall off at higher frequency while the -ImZ$_{Net}$ spectrum shows a peak at high frequency. We collected impedance spectra for our networks at a range of applied strains between 0 and 0.9%. These spectra were very similar to those collected at ε=0, only displaying significant deviations in the plateau region of ReZ$_{Net}$ and around the peak of -ImZ$_{Net}$, as illustrated in figure 3D.

Before analysing these spectra in detail we note that the real impedance at low frequency is equivalent to the DC resistance of the network.[38] From figure 3D we can see that increasing the applied strain results in a steady increase in the value of ReZ$_{Net}$ at low frequency, behavior which is equivalent to an increase in DC resistance of the network. We found the average value of ReZ$_{Net}$ (averaged between 125 and $1 \times 10^5$ rad/s) for each strain and used it to find the fractional resistance change for each strain (taking $\left([\text{Re}\,Z_{Net}(\varepsilon) - \text{Re}\,Z_{Net}(\varepsilon=0)] / \text{Re}\,Z_{Net}(\varepsilon=0)\right)_{\omega \to 0} = \Delta R / R_0$). As shown in figure 3E, we find a linear relationship between $\Delta R / R_0$ and strain, consistent with a gauge factor G$_{Net}$=3.4±0.2, reasonably consistent with the data in figure 2. The differences might be associated with a slight sample aging (the measurements were done on different days).

In general, impedance spectra are analysed by fitting them to equations associated with the appropriate equivalent circuit (figure 3A). Fitting the network impedance equations then yields values of R$_S$, R$_P$ and C$_P$. Although, spectra associated with equivalent circuits in figure 3A are often fitted with relatively simple equations, this is not always appropriate. If the values of R$_P$ & C$_P$ are controlled by microscopic elements (i.e. the resistance, R$_J$, and capacitance, C$_J$, of the individual inter-nanosheet junctions) which have a distribution of values, then both real and imaginary impedance spectra can be broadened somewhat.[49] Then, the real impedance and the imaginary impedance spectra are better modelled using Equation 4a and 4b respectively:[38, 49]

$$\text{Re}\,Z(\omega) = R_S + \frac{R_P + [1+(\omega R_P C_P)^n \cos(n\pi/2)]}{1 + 2(\omega R_P C_P)^n \cos(n\pi/2) + (\omega R_P C_P)^{2n}} \qquad (4a)$$

$$-\text{Im}\,Z(\omega) = \frac{R_P (\omega R_P C_P)^n \sin(n\pi/2)]}{1 + 2(\omega R_P C_P)^n \cos(n\pi/2) + (\omega R_P C_P)^{2n}} \qquad (4b)$$

where n is a parameter which describes the width of the distribution of values of the product of junction resistance and capacitance (R$_J$C$_J$) within the system (n=1 indicates an infinitely narrow distribution).[49]

Fitting the real impedance spectra yields values for $R_S$, $R_P$ and $C_P$ (as well as values of n≈0.83) for each strain while fitting the imaginary impedance spectra yields values for $R_P$ and $C_P$ (as well as values of n≈0.84). However, it is more instructive to convert these values to their equivalent microscopic counterparts, $R_{NS}$, $R_J$ and $C_J$, which represent the mean values of the resistance of individual nanosheets ($R_{NS}$) and junctions ($R_J$) as well as the mean capacitance associated with individual junctions ($C_J$).

Gabbett et al. showed that these conversions can be made using the following equations[38] (where we have applied the approximation that the nanosheet carrier density is high, as is usually the case for electrochemically exfoliated MoS$_2$)[16, 38]:

$$R_{NS} \approx \frac{W_{Ch} t_{Net}}{L_{Ch}} \frac{(1-P_{Net})}{2 t_{NS}} R_S \tag{5a}$$

$$R_J \approx \frac{W_{Ch} t_{Net}}{L_{Ch}} \frac{(1-P_{Net})}{2 t_{NS}} R_P \tag{5b}$$

$$C_J = R_P C_P / R_J \tag{5c}$$

Here $t_{Net}$=(19.6 ± 1.2) nm is the network thickness, $P_{Net}$ represents the network porosity (which we approximate to be zero in these networks) and $t_{NS}$=3.3 nm is the mean nanosheet thickness. In addition, $L_{Ch}$ and $W_{Ch}$ are the channel length and width. Both depend on applied strain, ε, as strain will elongate the Kapton substrate, leading in an increase in $L_{Ch}$ and a corresponding decrease in $W_{Ch}$ (the Poisson effect). To account for these changes, we used the strain-dependent values of $L_{Ch} = L_{Ch,0}(\varepsilon+1)$ and $w_{Ch} = w_{Ch,0}(1-\nu_y \varepsilon)$. Here $L_{Ch,0}$=100 μm and $W_{Ch,0}$=19.55 mm are the zero strain values while $\nu_y$ is the Poisson ratio of the Kapton substrate (0.39).[50] We assume the out of plane Poisson ratio is negligible ($\nu_z$≈0), such that the network thickness is constant.

We plot the resultant values of $R_{NS}$, $R_J$ and $C_J$ versus applied strain in figures 3 F-H. Shown in Figure 3F, are $R_{NS}$ values obtained from fitting the real impedance spectra. The zero strain nanosheet resistance was 138±3 kΩ, equivalent to a nanosheet conductivity of $\sigma_{NS}$~1000 S/m, slightly higher than previously measured.[38] This discrepancy is probably due to differences in doping state due to the influence of the substrate and the presence of trapped solvent. Notably, we find that there is no significant change in the nanosheet resistance for low applied strain i.e. $dR_{NS}/d\varepsilon \approx 0$. As MoS$_2$ nanosheets are known to have an intrinsic gauge factor of $G_{NS}$=-50,[33] this means that the strain within the nanosheets themselves must be very small.

This implies that, for these particular networks at least, there is no stress transferred to the individual nanosheets. This suggests that, in response to an applied strain, the nanosheets slide relative to each other.

We plot $R_J$ values obtained from fitting both real and imaginary impedance spectra versus strain in figure 3G. Here, although we find slightly different $R_J$ values from each fitting type, both curves follow the same trend and differ only by ~1% from their common mean. The mean value of $R_J$ at zero strain was 432±3 k$\Omega$, a value which is somewhat lower than the previously reported value of ~2 M$\Omega$ for similar networks.[38] This difference is almost certainly due to morphological variations associated with differences in film thickness and substrate. Both the ReZ$_{Net}$ and -ImZ$_{Net}$ data show a linear increase of $R_J$ with strain with very similar slopes, of 1000±180 and 1120±160 k$\Omega$ (per unit of absolute strain). This leads to a mean slope of $dR_J/d\varepsilon \approx 1060\pm170$ k$\Omega$. Dividing through by the junction resistance at $\varepsilon$=0 gives the mean fractional junction resistance change. This is equal to the mean gauge factor associated with an individual junction: $G_J$=2.5±0.4.

The junction capacitance, $C_J$, obtained from fitting both real and imaginary impedance spectra is plotted versus strain in figure 3H. Both values are similar, generally within a few percent of their common means. The capacitance shows very little dependence on strain, a fact that we will return to below.

DISCUSSION

We are now in a position to calculate the two contributions to the overall gauge factor, $G_{Dimensional}$ and $G_{Intrinsic}$, as well as their sum which is the overall gauge factor ($G_{Net}$). The former is just the first term in equation 3a: $G_{Dimensional} = (1+\nu_y+\nu_z)$. To calculate this, we note that $\nu_y$ is the Poisson ratio of the Kapton substrate, i.e. 0.39.[50] In addition, $\nu_z$ is the Poisson ratio of the nanosheet network which is expected to be very small, ranging from -0.1 to 0.1.[51-57] Thus, we expect $G_{Dimensional} \approx 1.4$. In addition, now that we have values for nanosheet and junction resistance at $\varepsilon$=0 ($R_{NS,0}$ and $R_{J,0}$) as well as $dR_{NS}/d\varepsilon$ and $dR_J/d\varepsilon$, we can calculate the second term in equation 3b which represents the intrinsic piezoresistance of the network. This works out as $G_{Intrinsic}$=1.9±0.3, similar in magnitude to $G_{Dimensional}$. The overall gauge factor is then expected to be $G_{Net}$=3.3±0.3, very close to our value measured from the low-frequency ReZ$_{Net}$ data (figure 3E).

We can understand the nature of the intrinsic piezoresistance as follows. The data in figure 3F shows no significant change in nanosheet resistance with applied strain. Because MoS$_2$ nanosheets are known to have a considerable negative piezoresistance, this implies the nanosheets do not deform in response to the external strain but rather slide past each other.

In the SI, we derive a simple model which shows that, under such sliding, the average junction area under strain is given by

$$A_J = A_{J,0}\left[1 - \frac{(1-f_{J,0})}{f_{J,0}}\varepsilon\right] \qquad (6a)$$

where $A_{J,0}$ is the average unstrained junction area and $f_{J,0} = A_{J,0}/l_{NS}^2$. The junction resistance is then $R_J = (RA)_J / A_J$, where $(RA)_J$ is a fundamental property of the junctions. This means that

$$R_J \approx \frac{(RA)_J}{A_{J,0}}\left[1 + \frac{(1-f_{J,0})}{f_{J,0}}\varepsilon\right] \qquad (6b)$$

where we use the approximation that for small x: $(1-x)^{-1} \approx 1+x$. This model implies a linear relationship between R$_J$ and strain as we find experimentally. It also makes the prediction that the ratio of the slope to intercept of a linear fit to the data should be equal to $(1-f_{J,0})/f_{J,0}$. From the fits, the average slope/intercept=2.45±0.4. this implies that $f_{J,0}$=0.29±0.3. However, we know that $f_{J,0} = A_{J,0}/l_{NS}^2$ and that A$_{J,0}$=0.33 ± 0.03 μm$^2$ and l$_{NS}$=1±0.03 μm, giving $f_{J,0}$ =0.33±0.04. This is excellent agreement with the value above supports our simple model.

We can also consider the junction capacitance:[38] $C_J = \varepsilon_r\varepsilon_0 A_J / l_J$, where $l_J$ is the spacing between nanosheets at the junction. We can use equation 6a for A$_J$:

$$C_J = \varepsilon_r\varepsilon_0 \frac{A_{J,0}}{l_J}\left[1 - \frac{(1-f_{J,0})}{f_{J,0}}\varepsilon\right] = C_{J,0}\left[1 - \frac{(1-f_{J,0})}{f_{J,0}}\varepsilon\right] \qquad (6c)$$

Taking the negative of the value of slope/intercept found above, we can plot the expected behaviour on the graph in figure 3F. We find very little change in capacitance is expected under strain with the plotted lines more of less consistent with the error bars on the data points. Again, this shows that our simple model of nanosheet sliding is consistent with all of our observations.

Taken together, this analysis implies that applying strain to such aligned networks of nanosheets results in no significant deformation of the nanosheets themselves but rather sliding of the nanosheets across each other. This results in junction areas and hence junction resistances that change linearly with strain. It also implies that the network thickness is strain independent, consistent with previous reports that the Poisson ratio of nanosheet networks is extremely small.

Our values of network gauge factor of ~3 are relatively small, implying a roughly 3% increase in network resistance on application of a 1% strain. However, it may be difficult to reduce this further in these systems. The dimensional component is determined by the Poisson ratio of the substrate, a factor over which the user may not have much control within a given technology. The intrinsic component is governed by the change in junction area with strain, a natural consequence of sliding. If sliding were suppressed, for example by crosslinking the nanosheets, this would allow strain to build up within the nanosheets. Because of the large intrinsic piezoresistance of $MoS_2$, this would result in significant changes in $R_{NS}$ with strain. In fact the best way to reduce $G_{Net}$ might be to induce corrugations or other local deformations in the unstrained nanosheets such that applied strain modified neither nanosheet nor junction dimensions.

CONCLUSION

This study investigates the piezoresistive phenomenon in networks of electrochemically exfoliated $MoS_2$ nanosheets. Highly aligned nanosheet networks were assembled at liquid-liquid interfaces and deposited on flexible Kapton substrates. These were characterised electromechanically, measuring the change in resistance with strain for linear and sawtooth strain profiles. Further insight into the piezoresistive phenomenon in the nanosheet network was achieved by carrying impedance spectroscopy under strain. In this way, the nanosheet resistance, junction resistance and junction capacitance could be directly measured as a function of strain. In these $MoS_2$-based networks, we found the nanosheet resistance to be invariant with strain while the junction resistance increased linearly with strain. A simple model was used to relate these findings to a linear decrease in junction area with strain. This suggests that, in response to strain, the nanosheets slide past each other. Combining our results with a simple piezoresistive model allowed us to calculate the contributions to the gauge factor from both geometric and intrinsic factors. Both contributions are small and of similar magnitude. This study represents a complete characterisation of the effects of strain on electrical properties of these networks. For example, once $R_{NS,0}$, $R_{J,0}$, $dR_{NS}/d\varepsilon$ and $dR_J/d\varepsilon$ are known, one can use existing models to calculate the strain dependence of parameter such as network conductivity and mobility.[38]  This work enhances our understanding of piezoresistance in nanonetworks and may provide some clues to enable the further reduction of gauge factor in such systems.

METHODS

*MoS$_2$ Exfoliation*

Electrochemical exfoliation of the layered MoS$_2$ crystal used an electrochemical setup consisting of a platinum foil (Alfa Aesar) anode and the bulk crystal as the cathode. The electrolyte consisted of 50 mL propylene carbonate with tetrapropylammonium (TPA) bromide (Sigma-Aldrich) at a concentration of 5 mg/mL. A potential difference of 8 V was applied for 30 mins to expand the layered crystal by ion intercalation. The expanded crystals were washed with DMF to remove any residual bromine and propylene carbonate before bath sonication in DMF with poly(vinylpyrrolidone) (MW ~ 40,000, 1 mg/mL) for 5 mins. The dispersion was then centrifuged (Hettich Mikro 220) at 500 RPM for 20 mins to remove unexfoliated materials.

*Size Selection*

The dispersion was then size selected by centrifuging the supernatant (top 90% of liquid) at 1,000 RPM for 1 h. The sediment, which contains the size selected nanosheets was resuspended in 2 mL DMF and centrifuged at 10,000 RPM for 1 h, this step was repeated twice, to remove the residual PVP. The sediment was then resuspended in 0.5 mL IPA and centrifuged at 10,000 RPM to remove the residual DMF. The sediment was then redispersed in ~ 0.5 mL of IPA, yielding an MoS$_2$ dispersion with a concentration of ~ 2.5 mg/mL, which was used to deposit films.

*Characterisation*

UV-Vis spectroscopy (Perkin Elmer, Lambda 1050) was used to characterise the nanosheet based ink. Drop cast films of MoS$_2$ ink on Si/SiO$_2$, annealed at 120 °C for 30 min, were analysed using Raman spectroscopy (WITec Raman Spectrometer, 523 nm). Electron microscope imagery of the film was carried out with the Zeiss Ultra SEM, with a 3 keV potential, working distance of 5 mm, with the secondary electron detector.

*Substrate Preparation*

Kapon substrates (DuPont) were cleaned using IPA. Compressed $N_2$ was used to dry the samples. Evaporated electrodes (Ti/Au (5 nm/95 nm)) were deposited using a Temescal evaporator (FC-2000). A shadow mask was used to define the electrode dimensions of $L_{Ch}$ = 100 µm, and $W_{Ch}$ = 19.55 mm, with contact pads to facilitate electrical contact after film deposition. Immediately before film deposition, substrates were treated for 3 minutes using a UV Ozone Cleaner (Ossila).

*Film Deposition*

To deposit films using the Langmuir-Schaefer method, a custom setup was used, as described in recent publications.[16, 58] The custom substrate holder, with substrate mounted, was placed in a glass beaker (250 mL) which was filled with DI water until the holder was submerged. Distilled n-Hexane (2 mL) was added to the top of the water, forming a liquid-liquid interface. The nanosheet ink was injected into the interface using a Pasteur pipette, until uniform coverage was observed by eye. The substrate was lifted through the liquid-liquid interface, transferring the nanosheet film onto the substrate. The substrate was dried at room temperature, before annealing in an Argon atmosphere at 120 °C for 2 h.

*Thickness Characterisation*

The film thickness of the network was determined using a Filmetrics Profilm3D® optical profiler in phase shift interferometry mode and a 50X Nikon objective lens. Thickness was calculated by taking the stap height from the top of the film and a scratch in the film, to the Si/$SiO_2$ substrate below using the histogram measurement protocol on Profilm Online. Si/$SiO_2$ was used as a substrate as it is more planar for accurately measuring optical interference as the technique is sensitive to bending of the substrate.

*Electromechanical Testing*

The samples were mounted in the clamps of a Zwick Z0.5 ProLine Tensile Tester (100 N Load Cell). The tester applied tensile stress to the mounted device. DC electrical characterisation was carried out by connecting the sample to a Keithley (KE2601) electrical source meter in a two-probe setup to measure resistance.

*Impedance Spectroscopy Measurements*

Impedance spectra were measured using the Keysight E4990E Analyser, (max frequency: 30 MHz), with the attached 16047E connector, to connect the sample's wires to the analyser. Short (<5 cm) wires were attached to the gold contact pads on the substrate with conductive silver paint (Ted Pella) to avoid high frequency inductive artifacts. Spectra were using precision speed 3 and an amplitude of 500 mV.

ACKNOWLEDGEMENTS: This project has received funding from the European Union's Horizon Europe research and innovation programme under grant agreement No 101129613 (HYPERSONIC). We acknowledge the Irish Research Council (GOIPG/2020/1051) and a Marie Skłodowska-Curie Individual Fellowship "MOVE" (grant number 101030735). We have also received support from the Science Foundation Ireland (SFI) funded centre AMBER (SFI/12/RC/2278) and availed of the facilities of the SFI-funded advanced microscopy laboratory (AML), additive research laboratory (ARL) and iCRAG labs.

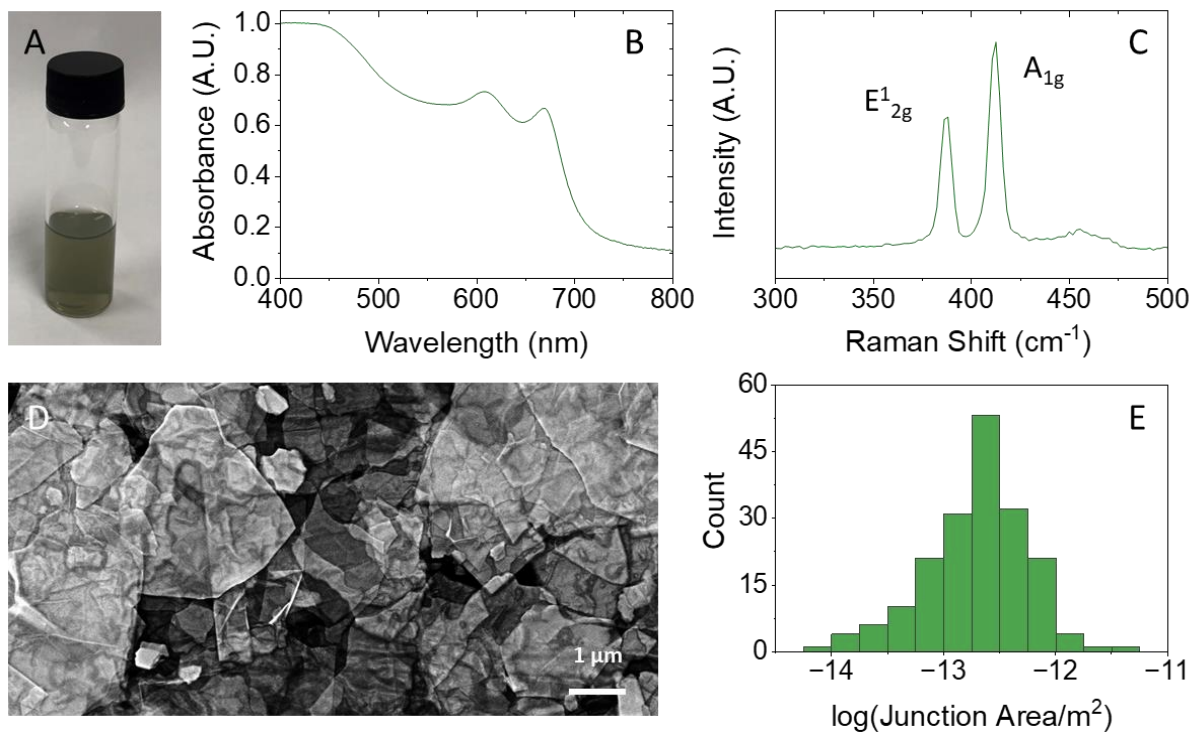

Figure 1: Ink and Network Characterisation A) Photograph of $MoS_2$ nanosheet ink suspended in IPA. B) Absorbance Spectrum of $MoS_2$ ink, showing the characteristic excitonic features. C) Raman Spectrum of a drop cast film of $MoS_2$ nanosheets. D) Top-down scanning electron microscope image of a deposited network of $MoS_2$ nanosheets on Kapton, showing the inter-

nanosheet junctions. E) Histogram of junction areas measured from scanning electron microscope images.

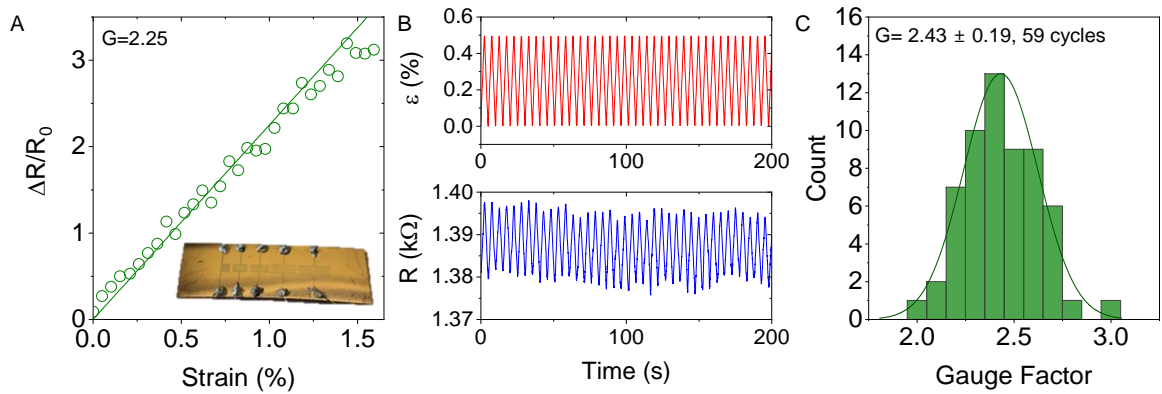

Figure 2: DC Piezoresistance A) Plot of the fractional change in resistance vs. strain, for a device strained linearly at 0.05 %/s. Inset photograph of the tested sample, showing electrodes and silver painted contacts. B) Cycling data for the device, stained to a maximum of 0.5%, with a triangular sawtooth strain profile, strain rate 0.2 %/s, with the corresponding DC resistance response. C) Histogram of the extracted gauge factor values by peak picking the maximum and minimum values from the cyclic profiles, showing a normal distribution.

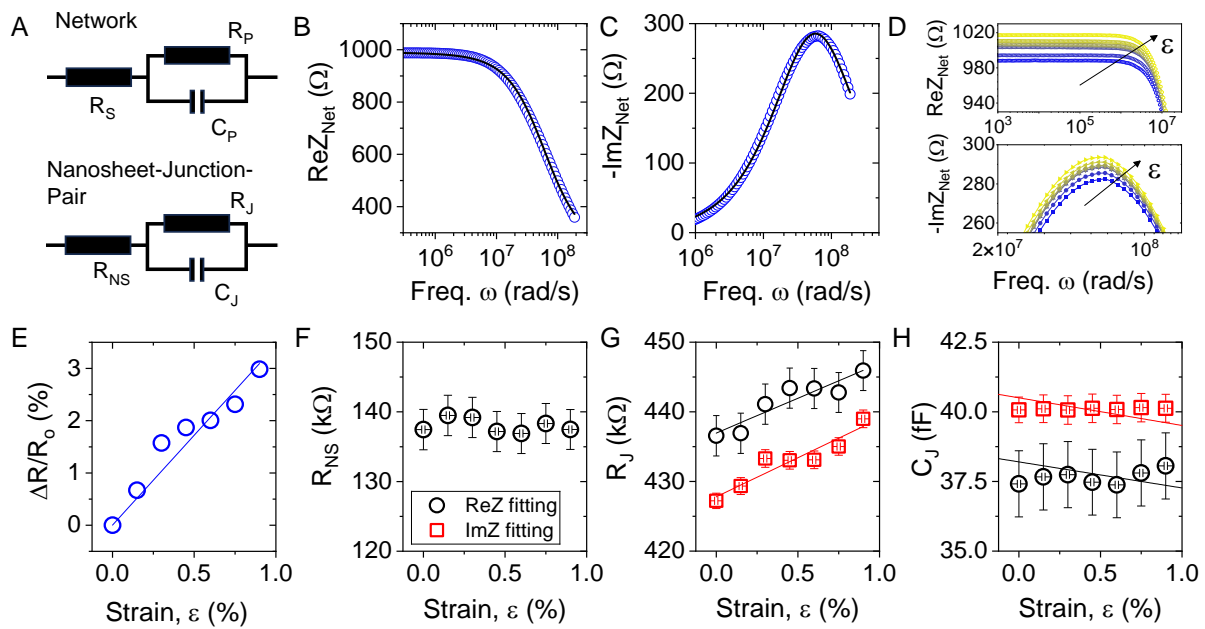

Figure 3: A) Equivalent circuit diagrams for: Top, the entire network; Bottom, an average nanosheet-junction pair. B-C) Real (B) and imaginary (C) network impedance vs frequency. In B and C, although the measurement was performed between 125 and $2\times10^8$ rad/s, only the upper end of this range is shown (the region between 125 and $10^6$ rad/s was featureless). D) The evolution of the real (top) and imaginary (bottom) impedance spectra as the applied strain is increased. E) Plot of the fractional change in the low frequency impedance vs strain, this is the AC equivalent of Figure 2A. F) Plot of nanosheet resistance with strain, extracted from fits of the real impedance spectra. G) Plot of junction resistance vs. strain, extracted from fits of real and imaginary impedance. H) Plot of junction capacitance vs. strain, extracted from fits of real and imaginary impedance. The legend in F also applies to G and H.